\documentclass[preprint]{aastex}
\newcommand{\be}{\begin{equation}}
\newcommand{\ee}{\end{equation}}
\newcommand{\bea}{\begin{eqnarray}}
\newcommand{\eea}{\end{eqnarray}}
\newcommand{\bvo}{{\bf v}_0}
\newcommand{\bvel}{{\bf v}}
\newcommand{\bvs}{{\bf v}_{\rm s}}
\newcommand{\nhat}{{\bf\hat{n}}}
\newcommand{\trs}{{\tilde{r}}_s}
\newcommand{\tr}{{\tilde{r}}_s}

\newcommand{\rhat}{{\bf\hat{r}}}
\newcommand{\bo}{\mbox{\boldmath $\mathbf{\omega}$}}
\newcommand{\bn}{\mbox{\boldmath $\mathbf{\nabla}$}}
\newcommand{\bth}{\mbox{\boldmath $\hat{\mathbf{\theta}}$}}
\newcommand{\bph}{\mbox{\boldmath $\hat{\mathbf{\phi}}$}}

\begin{document}

\title{The First Magnetic Fields}

\author{George Davies and Lawrence M. Widrow\altaffilmark{1}}
\affil{Department of Physics, Queen's University, Kingston, Ontario, Canada K7L 3N6}

\altaffiltext{1}{Visiting Professor, Department of Astronomy and Astrophysics, 
University of Chicago}

\begin{abstract}
We demonstrate that the Biermann battery mechanism for the creation of
large scale magnetic fields can arise in a simple model protogalaxy.
Analytic calculations and numerical simulations follow explicitly the
generation of vorticity (and hence magnetic field) at the
outward-moving shock that develops as the protogalactic perturbation
collapses.  Shear angular momentum then distorts this field into a
dipole-like configuration.  The magnitude of the field created in the
fully formed disk galaxy is estimated to be $10^{-17}\,{\rm Gauss}$,
approximately what is needed as a seed for the galactic dynamo.
\end{abstract}


\section{Introduction}

The origin of galactic magnetic fields has proved to be one of the
most challenging and stubborn problems in modern astrophysics (Rees
1987; Kronberg 1994 and references therein).  It is generally assumed
that galactic fields are generated and maintained by the dynamo action
of a differentially rotating disk galaxy.  However, a dynamo can only
amplify an existing field and so the question of galactic magnetic
fields splits naturally into two parts: creation of the field required
to seed the dynamo, and the nature of the dynamo itself (e.g.,
Zel'dovich, Ruzmaiken, \& Sokoloff 1983).

An early attempt to explain the origin of seed fields is due to
Harrison (1970, 1973) who showed that magnetic fields are created
during the radiation era if significant vorticity exists at that
epoch.  However, primordial vorticity in an expanding universe decays
with time (in contrast with the irrotational density perturbations
presumably responsible for structure formation).  Indeed, the absence
of significant vorticity prior to galaxy formation together with the
observation that vorticity is generic to galactic disks may provide an
important clue as to the origin of galactic magnetic fields.  Angular
momentum in galaxies is thought to arise from tidal torques among
neighboring protogalaxies (Hoyle 1949, Peebles 1969, and White 1984).
However, gravitational forces alone do not produce vorticity and
therefore its appearance must be due to `gasdynamical' processes such
as those that occur at oblique shocks.  These same processes also
produce magnetic fields by creating a so-called Biermann battery
(Biermann 1950) which drives electric currents in the plasma.  Since
oblique shocks are inevitable in collapsing gas clouds, the early
stages of structure formation provide a natural site for the
production of seed fields (Pudritz \& Silk 1989, Kulsrud et
al.\,1997).

There have, of course, been other attempts to explain the origin of
seed fields.  One possibility is that first fields were created in
stars and subsequently expelled into the interstellar medium
(Bisnovatyi-Kogan, Ruzmaiken, \& Syunyaev 1973).  Alternatively, seed
fields may have been created in the very early Universe through such
exotic phenomena as quantum field creation during inflation (Turner
and Widrow 1988, Ratra 1992), phase transitions (Quashnock, Loeb, \&
Spergel 1988, Vachaspati 1991, Field \& Carroll 1998), and topological
defects (Sicotte 1997).  By comparison, the protogalactic battery has
a certain simplicity and elegance since the necessary ingredients are
generic to models of galaxy formation.

Most discussions of galaxy formation ignore the influence of magnetic
fields (see, however Wasserman 1978 and Kim, Olinto, \& Rosner 1996).
To be sure, galactic magnetic fields play an important role in a
number of astrophysical processes such as star formation, cosmic ray
confinement and gasdynamics.  But while the energy density in galactic
magnetic fields is comparable to that in cosmic rays and in the
turbulent motion of the interstellar medium, it is considerably less
than the energy density associated with the global dynamics of a
galaxy.  This suggests that magnetic fields play a secondary role in
the formation and evolution of galaxies.  Nevertheless, models of
galaxy formation should be able to explain their origin.

In this work, we investigate the generation and early evolution of
vorticity and magnetic fields in the context of a detailed, albeit
highly idealized, model protogalaxy.  Recently, Kulsrud et al.\,(1997)
have attempted to follow the creation of protogalactic magnetic fields
in a cosmological hydrodynamic simulation of a cold dark matter
universe.  They find that seed fields can be produced on a variety of
cosmologically interesting scales.  Our work complements and enhances
theirs by considering a simpler system where both analytic and
numerical techniques can be employed.  In so doing, we are able to
understand these earlier results and some of their limitations in
terms of relatively simple physics and numerics.  Our work also makes
contact with various semi-analytic models of disk galaxy formation.

An outline of our scenario is as follows:

\begin{itemize}

\item We consider an isolated, nearly spherical, density perturbation
in an otherwise Einstein-de Sitter universe.  The cosmic fluid
consists of both collisional gas and collisionless dark matter.  We
assume that the scales of interest are significantly smaller than the
horizon so that a Newtonian treatment is adequate.

\item Each element in the fluid expands to a maximum or turnaround
radius (as measured from the center of the protogalaxy) before
collapsing with inner regions reaching turnaround first.  It is during
the early stages of collapse that an outward moving shock develops.
As infalling material passes through the shock, it is heated rapidly
and decelerated.

\item Vorticity is generated at the shock provided the velocity of the
infalling gas is not everywhere perpendicular to the shock surface.
We demonstrate this for an axisymmetric prolate protogalaxy where the
vorticity generated at the shock is in the azimuthal direction.

\item An external tidal torque applied to the protogalaxy generates
shear angular momentum which in turn couples to the vorticity in the
postshock region.  As an example we consider again the model
protogalaxy described above but now under the influence of a tidal
torque along one of its short axes.  The resultant shear field couples
to the vorticity generated at the shock to yield a large-scale
dipole-like vorticity field oriented along the direction of the tidal
torque, i.e., along what will ultimately be the spin axis of the
galaxy.  The concomitant magnetic field has the same geometry and
provides the seed field for subsequent dynamo action.

\end{itemize}

Most previous analyses of protogalactic field generation have sought
order of magnitude estimates for the seed field strength without
making direct contact to specific models of structure formation
(Pudritz \& Silk 1989; Lesch \& Chiba 1994).  Our results are in
agreement with these estimates and go one step further by providing a
clear and simple picture of the geometry of the seed field.  Our model
is in the spirit of the semi-analytic and numerical studies of disk
galaxy formation by Mestel 1963; Fall \& Efstathiou 1980; Katz \& Gunn
1991; Dalcanton, Spergel, \& Summers 1997 and others.  However, in
those works, angular momentum and vorticity are assumed {\it ab
initio} and they are therefore unable to shed light on the creation of
the first magnetic fields.  In contrast, our model explicitly follows
vorticity generation during the earliest stages of galaxy formation.

In Section 2 we review the vorticity-magnetic field connection, derive
an expression for the magnetic field generated at an oblique shock,
and apply the results to our model protogalaxy.  These analytic
calculations are enough to obtain an estimate for the magnitude of the
magnetic field as well as the general features of its geometry.  The
numerical simulations presented in Section 3 provide a check of these
results and also serve to illustrate some of the pitfalls inherent in
using simulations to study problems of this type.  These simulations
do not include angular momentum and so in Section 4, we present a
simple semi-analytic calculation for the postshock evolution of the
vorticity and magnetic field in the presence of shear angular
momentum.  We conclude, in Section 5, with a summary and discuss
directions for future work.

\section{Vorticity Generation in a Protogalaxy: Analytic Treatment}

\subsection{Vorticity-Magnetic Field Connection}

The evolution of a collisional fluid is described by the 
Euler equation
\be\label{euler}
\frac{\partial {\mathbf v}}{\partial t}+
\left ({\mathbf v}\cdot\bn\right ){\mathbf{v}}~=~
-\frac{1}{\rho}{\bn} p-\bn\psi
\ee
together with Poisson's equation for the gravitational potential
$\psi$, the continuity equation, an equation of state, and an energy 
equation.  Taking the curl of 
eq.\,\ref{euler} yields the following for the vorticity 
$\bo\equiv\bn\times {\mathbf v}$:
\be\label{omega}
\frac{\partial \bo}{\partial t}-
\bn\times\left ({\mathbf v}\times\bo\right )~=~
\frac{\bn\rho\times\bn p}{\rho^2}
\ee
Thus, while galaxies can acquire angular momentum through tidal
fields, vorticity arises through purely gasdynamical processes, namely
pressure and density gradients that are not colinear.

Biermann (1950) realized that a similar situation exists for magnetic
fields.  In the usual formulation of magnetohydrodynamics (MHD),
the evolution of a magnetic field is described by the equation
\be\label{mhd}
\frac{\partial {\mathbf B}}{\partial t}-
\bn\times\left ({\mathbf v}\times {\mathbf B}\right )
-\frac{c^2}{4\pi\sigma}\nabla^2{\mathbf B}=0
\ee
where $\sigma$ is the conductivity.
If ${\mathbf B}$ is initially zero, then it will be zero at all times.
However, the derivation of eq.\,\ref{mhd} assumes a form for Ohm's
law that is not strictly valid for an electron-ion fluid.  A careful 
treatment dictates that we include,
on the right hand side, the term
\be\label{extra}
{\mathbf\Gamma}=\frac{c}{e}
\frac{\bn n_e\times\bn p_e}{n_e^2}
\ee
where $n_e$ and $p_e$ are the number density and pressure of free
electrons.  Approximate local charge neutrality implies that
$n_e\simeq n_p\equiv \chi\rho/m_p$ where $n_p$
is the proton number density, $\chi$ is the ionization fraction, and
$m_p$ is the proton mass.  In addition, since the
electron temperature is expected to be approximately equal to the total gas
temperature, $p_e\simeq pn_e/\left (n_e+n_p\right )=p\chi/\left
(1+\chi\right )$.  We can therefore
write
\be\label{extra2}
{\mathbf\Gamma}=\alpha
\frac{\bn\rho\times\bn p}{\rho^2}
\ee
where $\alpha\equiv m_p c/e\left (1+\chi\right ) 1.05\times \simeq 10^{-4} 
\mbox{Gauss}\cdot\mbox{s}$ (Kulsrud et al.\,1997).

In the limit of vanishing diffusion, the equations for $\bo$ and
${\bf B}$ take identical forms. Together with the assumption that initially
both the vorticity and magnetic field are zero, we have the relation,
\be\label{omega2b}
{\mathbf B} = \alpha \bo \simeq 10^{-4} \bo
\ee
where the units of ${\mathbf B}$ and $\bo$ are Gauss and Hz
respectively.  The growth and evolution of the magnetic field therefore 
mirrors that of the vorticity up until the time when
the diffusive effects of viscosity and
conductivity become important\footnote{ See again, Kulsrud et
al.\,(1997) for a further discussion of this point.}.

\subsection{Shock Wave Preliminaries}

For a barytropic fluid, $p=p(\rho)$ and therefore ${\mathbf\Gamma}=0$.
However, at curved shocks, the equation of state is more complicated
($p=p(\rho,s)$ where $s$ is the entropy) reflecting the fact that bulk
kinetic energy can be converted to thermal energy.  We therefore
expect $\mathbf\Gamma\ne 0$ and hence the generation of both vorticity
and magnetic fields (Kulsrud et al.\,1997).

An ideal shock can be treated as a surface of discontinuity in the gas
flow.  By imposing certain jump conditions at this surface we can
derive a relationship between the velocity field of the gas in the
pre- and post-shock regions and hence an expression for the vorticity
generated at the shock.  In this way, we bypass eq.\,\ref{extra2} and
avoid dealing with the complicated gasdynamics that occurs inside the
shock
  
The standard shock wave jump conditions (e.g.,
Landau and Lifshitz 1997) consist of a set of three relations that
guarantee the conservation of mass, momentum, and energy across the
shock.  Consider a point on a shock surface with velocity $\bvs$ and
normal $\nhat$.  Let $\rho_0$, $\bvo$, $p_0$, and $h_0$ be the
density, velocity, pressure, and enthalpy in the preshock region and
$\rho$, $\bvel$, $p$, and $h$ be the corresponding quantities in
the postshock region.  The jump conditions are
\bea
\rho u&=&\rho_0 u_0 \\
p+\rho u^2&=&p_0+\rho_0 u^2_0 \\
h+ \frac{1}{2}u^2&=&h_0+ \frac{1}{2} u^2_0 
\eea
where $u_0$ and $u$ are the velocity components along $\nhat$
in the rest frame of the shock:
\be\label{udef}
u_0\equiv \left (\bvo-\bvs\right )\cdot\nhat~~~~~~~~~~
u\equiv \left (\bvel-\bvs\right )\cdot\nhat~.
\ee
In addition, we have that the component of the velocity 
tangent to the shock is continuous.

For the situation at hand, the preshock gas is relatively 
cold and we can therefore set $p_0\simeq 0\simeq h_0$.  The three
jump conditions are then easily combined to give
\be\label{newu}
u~=~\left (\frac{\gamma-1}{\gamma+1}\right )u_0
\ee
where $\gamma$ is the polytropic index of the infalling
gas (equal to $5/3$ for an ideal gas).  This allows us to express
the velocity in the postshock region in terms of the preshock velocity:
\be\label{newv}
\bvel ~=~\bvo +(f(\gamma)-1)u_0\nhat
\ee
where $f(\gamma)\equiv (\gamma-1)/(\gamma+1)$.
 
\subsection{Vorticity in the Postshock Region}

The velocity field in the postshock region is determined not only from
the initial velocity field and the geometry of the shock (through
eq.\,\ref{newv}) but also from the evolution of the gas once it has
passed through the shock.  All of this is incorporated
into numerical simulations discussed in the next section.  Here
we present an analytic model for an idealized protogalaxy.  The key
simplification is to ignore the evolution of the gas in the postshock
region.  Our picture is that the shock wave sweeps
through the gas transforming the velocity field from ${\mathbf v}_0$
to ${\mathbf v}$ according to eq.\,\ref{newv} where $u_0$ and $\nhat$
are evaluated at each point at the instant when the shock passes
through.  In this way, $u_0$ and $\nhat$ can be treated as functions
of position and we can then calculate the vorticity by taking the curl
of ${\mathbf v}$.  As discussed above, we expect ${\mathbf v}_0$ to be
curl-free so that
\be\label{vorticity_ps} 
\bo=\left
(f(\gamma)-1\right ) \bn\times \left (u_0\nhat\right ) 
\ee 

To proceed further we require a specific model protogalaxy.  As a
starting point, we consider the spherical infall model (e.g., Gunn
1975, Gott 1977, Fillmore \& Goldreich 1984, Bertschinger 1985, Ryden
\& Gunn 1987, Ryden 1988) wherein matter is divided into spherical
shells which expand to a maximum or turnaround radius and then
collapse toward the center.  In the case of collisional matter (e.g.,
Bertschinger 1985) infalling shells are decelerated and heated as they
pass through an outward moving shock.

Current theories of structure formation present a far more complicated
picture than that represented by the spherical infall model.  In
particular, structure formation is believed to proceed hierarchically,
with subgalactic objects forming first, and then coalescing to form
galaxies and clusters.  For our purposes, the key deficiency of the
spherical infall model is its restriction to spherical symmetry, since
this precludes vorticity generation.  Moreover, a spherically
symmetric protogalaxy cannot acquire angular momentum through tidal
torques.  Of course, there is no reason to expect protogalaxies to be
spherically symmetric.  Indeed, the halos found in collisionless
N-body simulations are generally triaxial, with prolate shapes favored
slightly over oblate ones.  Furthermore, the angular momentum vector
for these systems is generally aligned with the short axis of the halo
(Carlberg \& Dubinski 1991; Warren et al.\,1992).  These results
motivate us to consider a simple model protogalaxy that forms from an
axisymmetric, prolate density perturbation.  An external tidal torque,
applied perpendicular to the symmetry axis of the perturbation,
generates shear angular momentum.  In the spirit of the spherical
infall model, we assume that the perturbation is smooth and
featureless with a density profile that decreases with radius.  The
protogalaxy therefore forms from the inside out.

For the moment, we ignore tidal torques.  The perturbation will
therefore evolve into an axisymmetric protogalaxy.  Consider a
spherical coordinate system, $(r,\theta,\phi)$, with polar axis
oriented along the symmetry axis of the protogalaxy.  The density and
pressure gradients that occur, for example, at an outward moving
shock, will be in the $\rhat$ and $\bth$ directions which imply that
any vorticity generated will be along the $\bph$ direction.  Moreover,
by symmetry the vorticity above and below the equatorial plane will be
in opposite directions.  This is not surprising since, in the absence
of tidal fields, the net circulation of the system must be zero.

It is instructive to consider a simple ansatz for the evolving
protogalaxy.  Specifically, we assume that isodensity contours for the
gas are concentric spheroids, i.e., $\rho=\rho(\tr)$ where
$\tr^2=r^2\left (\sin^2{\theta}+\cos^2{\theta}/q^2\right )$.  $q$ is
the flattening parameter which, in general depends on time and radius.
For simplicity, we will ignore this complication (we focus
on a small region in the neighborhood of the shock) and further assume that the
deviation from spherical symmetry is small ($|q-1|\ll 1$).  The shock
surface is described by the equation 
\be\label{shocksurface}
\trs^2(t)~=~r^2\left ( \sin^2{\theta}+\cos^2{\theta}/q^2\right ) \ee
and the normal to this surface is, to first order in $(q-1)$,
given by 
\be\label{shocknormal} \hat{\bf
n}=\hat{\bf r}+2\left (q-1\right )\sin{\theta}\cos{\theta} \bth 
\ee
Thus, one contribution to $\bo$ will be of the form $2\left
(f(\gamma)-1\right )\left (1-q\right ) \left (u_0/\trs\right
)\sin{\theta}\cos{\theta} \bph$.   There is a second contribution that
is proportional to $\bn u_0\times\nhat$ which has a similar
form\footnote{To evaluate this term, we
require an ansatz for the preshock gas flow, $\bvo$.  Since this is
assumed to be irrotational, it can be written as the gradient of a
scalar function.  A reasonable ansatz (akin to the Zel'dovich
approximation) is $\bvo\propto \bn\psi$.  In any case, we expect that
$\bvo\cdot\bth/\bvo\cdot\rhat = O(|q-1|)$ and likewise for $\bvs$ so
that $u_0=v_{0r}-v_{sr}+O((q-1)^2)$.  Moreover, $u_0$ should have the
form $u_0(r,\theta)=u_0^1+u_0^2(q-1)\cos^2(\theta)$ where
$u_0^1$ and $u_0^2$ are functions of $r$ which depend on the details
of the model.}
and we can therefore write 
\be\label{omega_explicit} 
\bo\propto \left (q-1\right ) \left
(u_0/\trs\right )\sin{\theta}\cos{\theta} \bph 
\ee
The magnitude of the vorticity is therefore set by the velocity of
the infalling gas (in the rest frame of the shock)
divided by the shock
radius.  This is roughly equal to the reciprocal of the turnaround
time, $T_{\rm ta}$, a result we might have anticipated from
dimensional analysis.  
As a specific example, consider Bertschinger's (1985) solution for
secondary infall of collisional matter onto an already collapsed
overdensity.  In this self-similar model, the turnaround radius at
time $t$ is $r_{\rm ta}(t)\propto t^{8/9}$ and the radius of the shock
is $r_s(t)=\lambda_s r_{\rm ta}(t)$ where $\lambda_s$ is a constant
$\simeq 0.33$ for $\gamma=5/3$.  Likewise the velocity of the gas,
immediately before passing through the shock, is given by
$v_0=-Vr_{\rm ta}/t$ where $V\simeq 1.47$ for $\gamma=5/3$.
This implies that $u_0/r_s\simeq 1.7/T_{\rm ta}$.

For a galaxy-sized object, $T_{\rm ta}\sim 10^{16}\,{\rm s}$ and
therefore $\omega_\phi\sim 10^{-16}\,{\rm s}^{-1}$.  This is roughly a
factor of $10$ less than the local value of the vorticity in the
Milky Way as determined from the Oort constants (e.g., Binney \&
Merrifield 1998), a reasonable result given that the vorticity will be
amplified during the formation of the disk itself.  The strength of the
corresponding magnetic field is $\sim 10^{-20}\,{\rm Gauss}$.
We will return to this result in the next section.

\section{Numerical Simulations}

In this section we present the results of numerical simulations that
are designed to test and augment the analytic model described above.
The simulations follow the evolution of an isolated axisymmetric
density perturbation in an otherwise flat (Einstein-de Sitter)
universe.  The cosmic fluid consists of dark matter and gas in a 10:1
ratio.  The simulations are performed using HYDRA (Couchman, Thomas,
\& Pearce 1995): Gravitational forces are calculated with an adaptive
particle-particle particle-mesh (${\rm AP^3M}$) algorithm while
gasdynamics is treated using smooth particle hydrodynamics (SPH).
Simulations are run with $32^3$ particles of each species.

The initial density profile has the form $\rho(r,\theta)=
\rho_b(t)\left (1+\delta(\tilde{r})\right )$ where $\rho_b(t)$ is the
background density for an Einstein-de Sitter Universe, 
$\tilde{r}=r\left (\sin^2{\theta}+\cos^2{\theta}/q^2\right )^{1/2}$, and 
\be\label{init} 
\delta(r)=\left\{\begin{array}{ll}
\delta_0\left (1-\frac{\alpha}{\alpha+2}\left (\frac{r}{r_c}\right )^2\right )
&\mbox{$r<r_c$}\\
\delta_0\left (\frac{2}{\alpha+2}\left (\frac{r}{r_c}\right )^\alpha\right )
&\mbox{$r_c<r<R$}\end{array}\right.
\ee 
To set up initial conditions, we begin with an interlaced lattice of
gas and dark matter particles.  Those particles a distance $R$ from a
chosen center are discarded and the ones that remain are displaced
from their original lattice sites so as to achieve the desired density
profile.  Velocities are then assigned according to the Zel'dovich
approximation.  In the simulations presented here, $\delta_0=1$,
$r_c/R=0.4$, $\alpha=2$, and for the prolate runs, $q=1.5$.
With this choice of parameters, the mean density enhancement is
$\bar{\delta}=0.377$.

The simulation units are such that the total mass $M=1$ and
Newton's constant $G=0.0194$.  Neither cooling nor star formation
are included in the simulations and so there is some freedom in choosing
units for dimensional quantities.  In conventional models of
structure formation, such as the Cold Dark Matter scenario (CDM) and its
variants, $\bar{\delta}$ can be identified with the rms mass
fluctuation on a scale $R$:
\be\label{sigmam}
\sigma_M(t)\equiv\langle\left (\Delta M/M\right )^2\rangle^{1/2}=
\left (1+z\right )^{-1}
\left (\int\frac{k^2 dk}{2\pi^2} P(k)W^2(kR)\right )^{1/2}
\ee
where $M\simeq 1.2\times 10^{12}h^2M_\odot\left (R/{\rm Mpc}\right
)^3$ is the total mass in a sphere of radius $R$, $P(k)$ is the linear
power spectrum for the model, $h$ is the present value of the Hubble
parameter in units of $100\,{\rm km\,s^{-1}\,Mpc^{-1}}$, $z$ is the
redshift and $W(x) = 3\left ( \sin{x} - x\cos{x}\right )/x^3$ is the
top hat window function.  Thus, by setting $\sigma_M=0.377$, we can
determine the initial redshift for the simulation, $z_i$, as a
function of $M$.  It is then straightforward to relate simulation
units to physical units.  This is done, for three representative
masses, in Table 1 where, in computing $\sigma_M$, we have assumed a
spatially flat CDM universe with $h=0.7$, $\Omega_B=0.05$, and COBE
normalization ($\sigma_8\simeq 1.7$).  The transfer function was calculated
using the fitting formula of Eisenstein and Hu (1999).

Figure 1 presents the phase space particle distribution (radius $r$
vs. radial velocity $v_r$) for the spherical run.  We see that the
turnaround radius, $r_{\rm ta}$, increases with time.  For $r>r_{\rm
ta}$, the gas and dark matter particles evolve as a single fluid.  For
$r<r_{\rm ta}$, the dark matter particles exhibit multiple phase space
streams that are characteristic of collisionless infall (cf Figure 10
of Fillmore \& Goldreich 1984 and Figure 6 of Bertschinger 1985).
Note however that at late times, these streams, especially the outward
moving ones, become rather chaotic.  This is a result of an
instability in the spherical infall model first described by Henriksen
\& Widrow (1997).  In contrast, the gas particles are decelerated with
$v_r\to 0$ for $r\to 0$.  The presence of an outward moving shock is
clearly seen in Figure 2 where we plot the temperature $T$ as a
function of $r$ for frames (b) and (d) of Figure 1.

We next turn to vorticity.  In SPH one determines the evolution of a
fluid system by following the motion of fiducial particles which are
labeled with local kinematic and thermodynamic quantities (for a
review, see Monaghan 1992 and references therein).  Any function
$f({\bf r})$ of these quantities may be approximated by the following
summation: 
\be\label{sphA} f({\bf r})=\sum_{b=1}^{N}\frac{m_b}{\rho_b}W({\bf
r}-{\bf r}_b;R)f_b 
\ee 
where $f_b$ is the value of $f({\bf r})$ for the $b$'th
particle and 
\be\label{rho} \rho_b=\sum_{b=1}^{N}m_b W({\bf r}-{\bf r}_b;R)
\ee 
In these expressions, $W$ is a user-supplied window function with
characteristic radius $R$.  This measurement process introduces an
error which can be minimized (but not eliminated) by an appropriate
choice of $R$.  Quantities that involve gradients require a bit more
care and the measurement prescription is not always unique.  Following
Monaghan (1992) the vorticity is determined as follows:
\be\label{vorticitysph} 
\bo_a=\left (\bn\times{\mathbf v}\right )_a=
\frac{1}{\rho_a}\sum_{b=1}^{N} m_b{\mathbf v}_{ba}\times \bn_a W_{ab}
\ee 
where ${\mathbf v}_{ba}\equiv {\bf v}_b-{\bf v}_a$, $W_{ab}\equiv
W({\bf r}_a-{\bf r_b};R)$ and $\bn_a W_{ab}$ denotes the gradient of
$W_{ab}$ with respect to ${\bf r}_a$.  Following the usual practice, 
we choose the ${\bf r}_a$ to be the positions of the particles themselves
though in principle ${\bf r}_a$ can be taken to be at any point in the
simulation volume.

In addition to the measurement error discussed above, there is an
error associated with the integration of particle orbits.  Moreover,
the ``boxy'' nature of the particle distribution, an artifact of the
initial conditions, will lead to vorticity generation even with $q=1$
since we do not have true spherical symmetry.

As a diagnostic test of these potential difficulties we determine the
vorticity field in the spherical run discussed above.  The result is
shown in Figure 3 where we plot the $(r,\theta,\phi)$ components of
the measured $\bo$ as a function of $r$ for frame (b) of Figure 1.
The large $\theta$ and $\phi$ components imply that there are angular
gradients in $v_r$ and/or radial gradients in $v_\theta$ and $v_\phi$.
For exact spherical symmetry and properly treated gasdynamics, the
velocity fields should be purely radial and the only gradients in the
$r$ direction.  $\omega_\theta$ and $\omega_\phi$ represent the first
terms that arise when spherical symmetry is broken.  In contrast, a
nonzero $\omega_r$ requires angular gradients in the tangential
velocity field and is therefore second order in small quantities and
so it is not surprising that the measured $\omega_r$ is the smallest
of the three components.  As expected, the amplitudes of
$\omega_\theta$ and $\omega_\phi$ decrease with increasing particle
number, roughly as $N^{-1/3}$.

As mentioned above, errors are introduced into the vorticity
calculation simply because we are attempting to determine a continuous
field from information at discrete and irregularly spaced points.  In
order to quantify this aspect of the problem, we calculate the
vorticity for a distribution of particles with the same positions as
those used to generate Figure 3 but with velocities chosen by hand to
reproduce a prescribed velocity field.  Equation\,\ref{vorticitysph}
is then used to determine the ``measured'' vorticity field.  For this
experiment, we assume a prescribed velocity field of the form
$v_z=\left (x^2+y^2\right )^{1/2}$ which implies a constant vorticity
field, $\bo=\bph$.  The measured field, shown in Figure 4, indicates
that for most particles, the SPH prescription does a good job of
calculating the vorticity.  However, for a subset of particles, errors
of order unity are introduced.

The difficulties inherent in following the generation and evolution of
vorticity in hydrodynamic simulations is apparent in the simulations
of Kulsrud et al.\,(1997).  They determine the magnetic field by
solving eq.\,\ref{mhd} with the additional term eq.\,\ref{extra2}
included on the right hand side.  As a check, they compare the result
with the vorticity (scaled by the appropriate constant) and find
discrepancies of order unity (cf. their Figure 4).

We next calculated the vorticity for the prolate protogalaxy
simulation (Figure 5).  As expected, there is now significant
vorticity generated at the shock, primarily in the azimuthal
direction.  The vorticity in the $r$ and $\theta$ directions is again
an artifact of the simulation.  The rms of $\omega_\phi$ is a factor
of $5$ greater than that of $\omega_\theta$ and $\omega_r$.  This can
be viewed, in some sense, as a measure of the signal-to-noise of the
simulation.

In Figure 6, $\omega_\phi$ is plotted as a function of $\theta$.  The
expected antisymmetry about the equatorial plane ($\theta=\pi/2$) is
readily apparent.  In particular, the vorticity near the poles
$(\theta=0$ and $\pi$) vanishes.  However, we also find that there is
vorticity generated with the ``wrong sign'', i.e., $\omega_\phi<0$ for
$0<\theta<\pi/2$ and $\omega_\phi>0$ for $\pi/2<\theta<\pi$.  This can
be understood as follows: When the gas flows through the shock, it is
refracted away from the symmetry axis and toward the equatorial plane.
This leads to a region of high pressure and density in the equatorial
plane forcing the gas to move out along the symmetry axis and creating
a region of vorticity with the opposite sign.  This is illustrated in 
Figures 7a and 7b where we show the velocity field of the particles
in the simulation.

For a system mass of $7\times 10^{11}M_\odot$, corresponding to a
spiral galaxy roughly the size of the Milky Way.  The magnitude of the
vorticity is $\simeq 10^{-15}\,{\rm s}^{-1}$ in good agreement with
our earlier estimate.  The magnitude of the corresponding magnetic
field is $\simeq 10^{-19}\,{\rm G}$.  The protogalaxy at these early
stages is roughly $25\,{\rm kpc}$ in size whereas the actual disk will
have a radius $\sim 10\,{\rm kpc}$ and a thickness $\sim 1\,{\rm
kpc}$.  Contraction of the protogalaxy in the plane of the disk will
therefore amplify the seed field by a factor of $(25/10)^2\simeq
6$ while collapse perpendicular to this plane will amplify the field
by a factor $\sim 25$ (Lesch \& Chiba 1995).  We therefore expect a
field strength in the fully assembled disk galaxy of $1.5\times
10^{-17}\,{\rm Gauss}$.  This is approximately what is required to seed
the galactic dynamo.

An alternative scenario is to generate the first magnetic fields 
in $10^6\,M_\odot$ objects.  The seed fields are approximately two orders
of magnitude larger (see Table 1).  More importantly, the dynamical
time for these systems is significantly shorter.  It may therefore be
possible for dynamo action to amplify fields on these scales before
the disk is assembled.

\section{Post-Shock Evolution}

In the axisymmetric model described above, the vorticity and magnetic
field generated at the shock are in the azimuthal direction and are
antisymmetric about the equatorial plane.  Mixing of gas from above
and below this plane will lead to a rapid decrease in the vorticity, a
reflection of the fact that angular momentum has not been included.
The evolution of the magnetic field involves recombination and is
therefore more complicated.  It is however clear that no large-scale
coherent field will survive without the addition of angular momentum.
As discussed above, shear angular momentum is generated by tidal
interactions with neighboring protogalaxies and is typically oriented
along one of the short axes of the protogalaxy.  It is the action of
the shear field on the vorticity and magnetic fields that leads
ultimately to a dipole configuration for these fields.  This process
can be illustrated by the following simple calculation.  A set of
particles are used to represent fluid elements labeled by their
position, velocity, velocity gradient, and magnetic field.  We assume
force-free evolution so that each particle evolves independently
according to the following (Cartesian coordinate) equations:
\be
\frac{dx_i}{dt}~=~v_i~~~~~~~~~~\frac{dv_i}{dt}~=~0 
\ee
\be
\frac{d\partial_i v_j}{dt}~=~\left (\partial_i
v_k\right ) \left (\partial_k v_j\right )~~~~~~~~~~
\frac{dB_i}{dt}~=~B_j\partial_jv_i-B_i\partial_j v_j 
\ee 
Initially, the magnetic field is in the azimuthal direction and is
antisymmetric about the equatorial (xz) plane (Figure 8a).  The
prescribed velocity field includes shear angular momentum about the
$z$-axis and an inward radial flow.  The latter is meant to model the
continual contraction of the fluid under the influence of gravity.
After a short period of time, these field lines are sheared into a
dipole configuration (Figure 8b).

\section{Conclusions}

In summary, we have presented a detailed investigation of magnetic
field generation during the collapse of a protogalactic density
perturbation.  The first fields appear, via the Biermann battery
effect, in the region of the outward moving shock that develops in the
collapsing protogalaxy.  Shear angular momentum is then able to
reconfigure the field into a dipole pattern oriented along the spin
axis of the protogalaxy.  The predicted field strength, once the disk
has formed, is estimated to be $10^{-17}\,{\rm Gauss}$.  With a seed
field of this magnitude, dynamo action can create microgauss fields by
the current epoch.  The magnitude of the associated vorticity at the
time of disk formation is roughly equal to its present day value.

Virtually all galactic dynamo models assume azimuthal symmetry with
respect to the spin axis of the disk.  The dynamo equations in these
models possess an invariance with respect to reflections about the
equatorial plane and therefore the solutions can be divided into two
groups: odd modes which consist of a dipole-like poloidal field
together with an antisymmetric toroidal field and even modes which
consist of quadrupole-like poloidal fields together with symmetric
toroidal fields.  In general, even modes are favored (faster growing)
when the fields are confined to a disk, while odd modes are favored in
more spherical configurations (e.g., Ruzmaikin, Shukurov, \& Sokoloff
1988).  Observations would seem to indicate that both types of
configurations are present in nature: The fields in the inner regions
of the Milky Way, for example, appear to be predominantly
antisymmetric with respect to the disk plane (Han et al.\,1997) while
those in M31 are evidently symmetric (Han, Beck, \& Berkhuijsen 1998).

Our analysis suggests that dipole-like seed fields are favored (see,
also Krause \& Beck 1998).  However, in a more realistic model, based
on hierarchical clustering, we expect both dipolar and quadupolar
fields to be produced.

If galactic magnetic fields have their origin in the Biermann battery
effect operating in protogalactic shocks, it should be possible to
follow the formation of a disk galaxy from primordial density
perturbation, with ${\bf B}=0$, to mature galaxy with a microgauss
field.  Our analysis and simulations have taken the first step in this
ambitious program.  Future work will include cosmological tidal fields
as well as small scale perturbations.  In addition, the magnetic field
will have to be treated explicitly since the correspondence with
vorticity is ultimately lost.

\acknowledgments{We wish to thank R.\,Henriksen, H.\,Couchman, and,
L. Chamandy for useful discussions.  LMW acknowledges the hospitality
of The University of Chicago during a sabbatical stay and GD
acknowledges the hospitality of the Canadian Institute for Theoretical
Astrophysics.  This work was supported in part by a grant from the
Natural Sciences and Engineering Research Council of Canada.}

{}

\newpage

\begin{deluxetable}{cccc}
\tabletypesize{\scriptsize}
\tablecaption{Physical units for numerical simulations}
\tablewidth{0pt}
\startdata
$[M]~M_\odot$   		&   $10^6$  & $10^9$  & $7\times 10^{11}$ \\
$z_i$ \tablenotemark{a} 	&   $120$   & $63$    & $25$              \\ 
$z_f$ \tablenotemark{b}		&   $43$    & $22$    & $8.4$	          \\
$[L]~(\rm kpc)$			&   $5.9$   & $11$    & $24$              \\
$[V]~(\rm km\,s^{-1})$		&   $14$    & $100$   &	$560$             \\ 
$[\omega]~(10^{-16}\rm s^{-1})$	&   $75$    & $29$    & $7.5$             \\ 
$[B]~(10^{-20}\,{\rm G})$	&   $79$    & $31$    & $7.9$             \\	
\enddata

\tablenotetext{a}{Redshift at the start of the simulation}
\tablenotetext{a}{Redshift corresponding to panel (d) of Figure 1}

\end{deluxetable}

\begin{figure}
\plotone{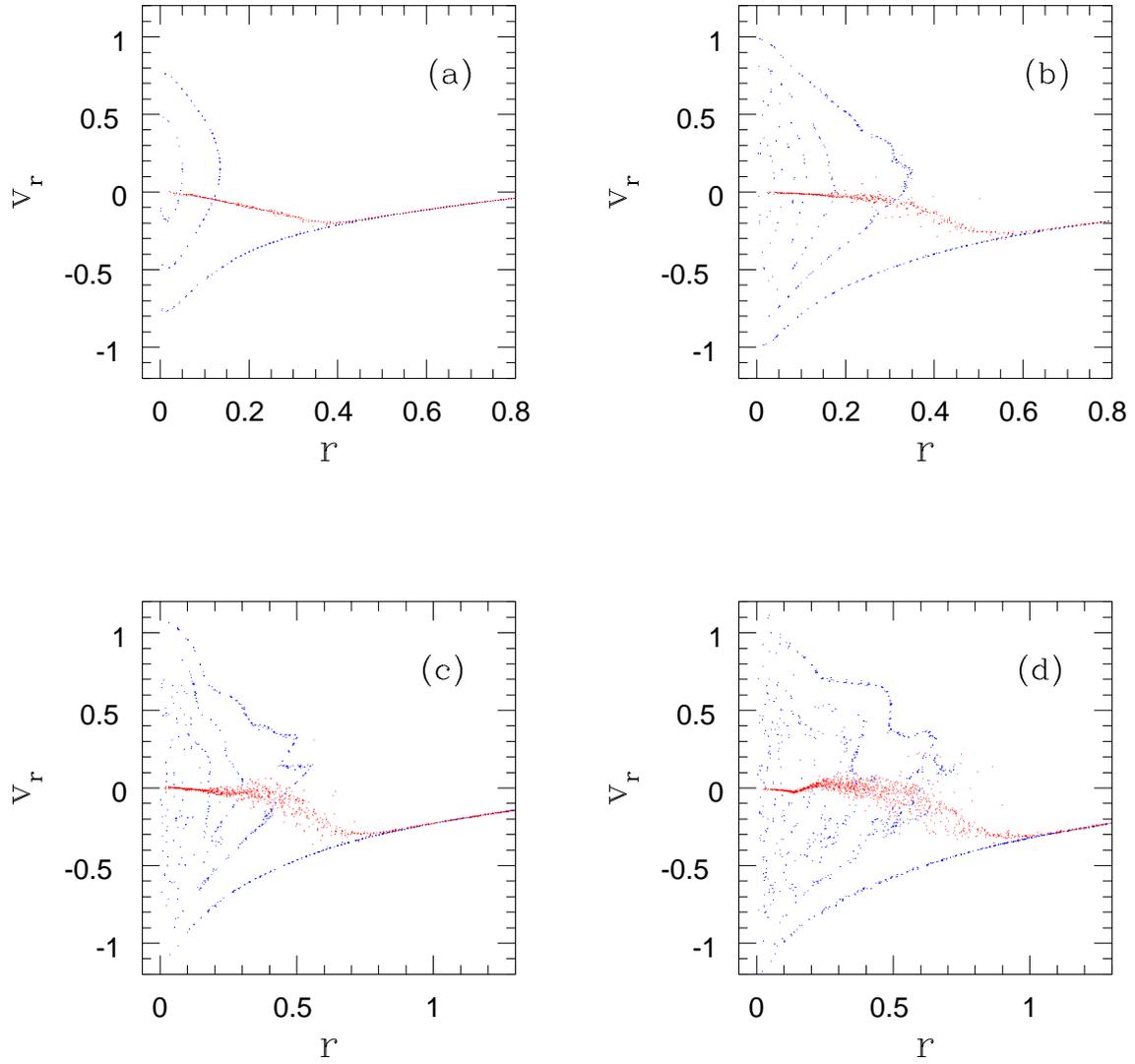}
\caption{Phase space (radius $r$ vs. radial
velocity $v_r$) distribution of particles in the spherical run.
The blue points represent dark matter particles while the red points
represent gas particles.}
\end{figure}

\begin{figure}
\plotone{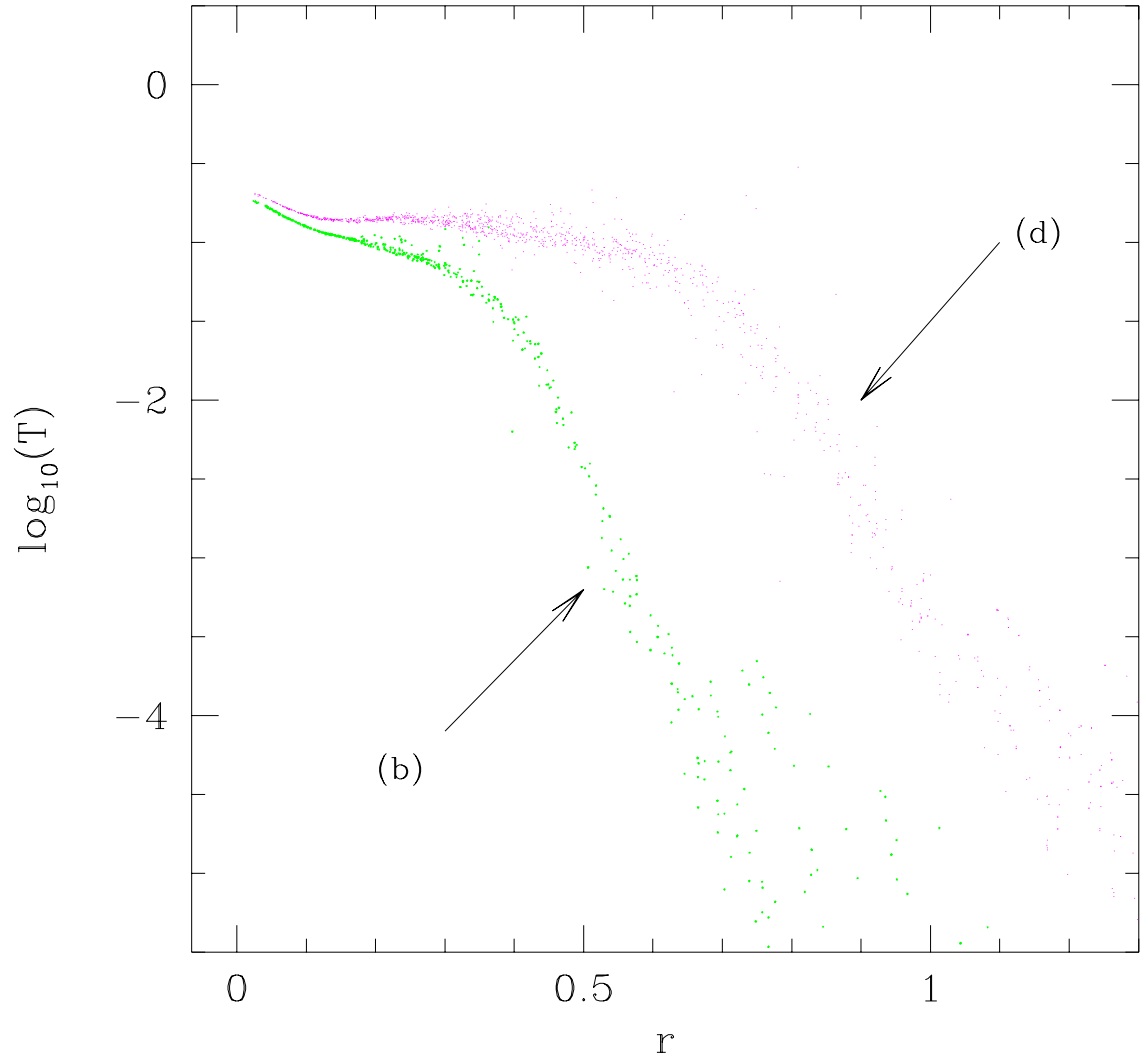}
\figcaption{Temperature as a function of radius
corresponding to frames (b) and (d) of figure 1}
\end{figure}

\begin{figure}
\plotone{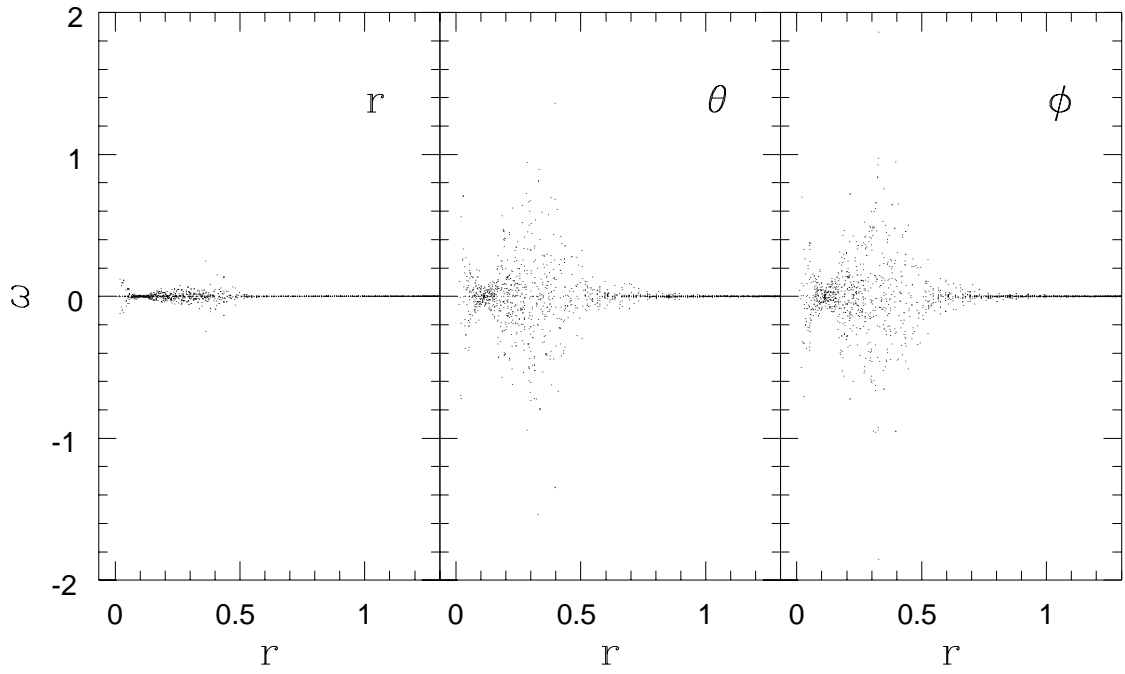}
\caption{$(r,\theta,\phi)$ components of $\bo$
for frame (b) of Figure 1.}
\end{figure}

\begin{figure}
\plotone{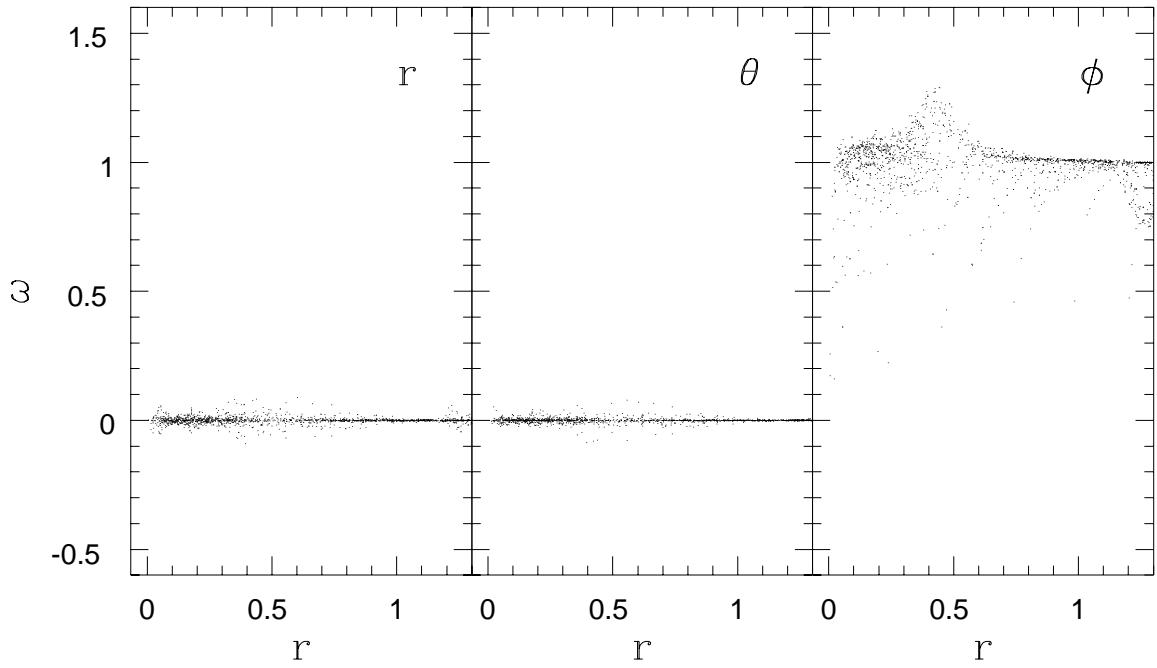}
\caption{Same as Figure 3 with a prescribed velocity
field corresponding to $\bo=\bph$.}
\end{figure}

\begin{figure}
\plotone{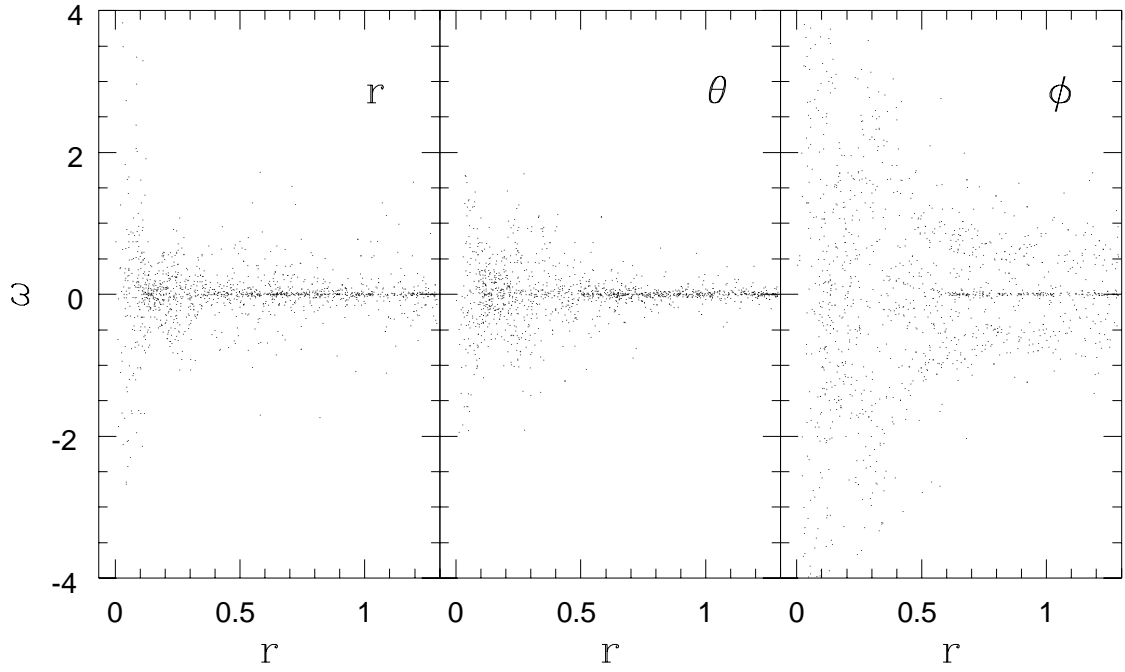}
\caption{Same as Figure 3 but for the prolate
protogalaxy ($q=1.5$).}
\end{figure}

\begin{figure}
\plotone{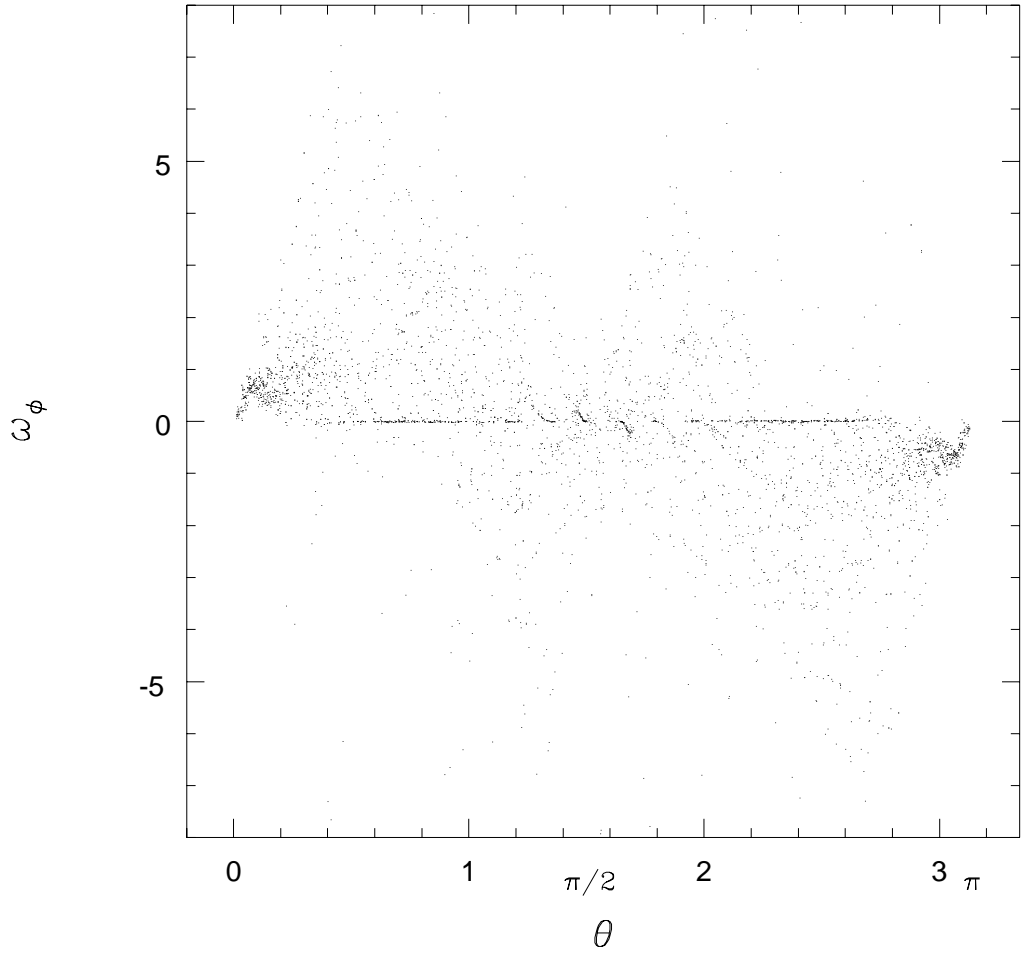}
\caption{$\omega_\phi$ as a function of $\theta$.}
\end{figure}

\begin{figure}
\epsscale{1.2}
\plottwo{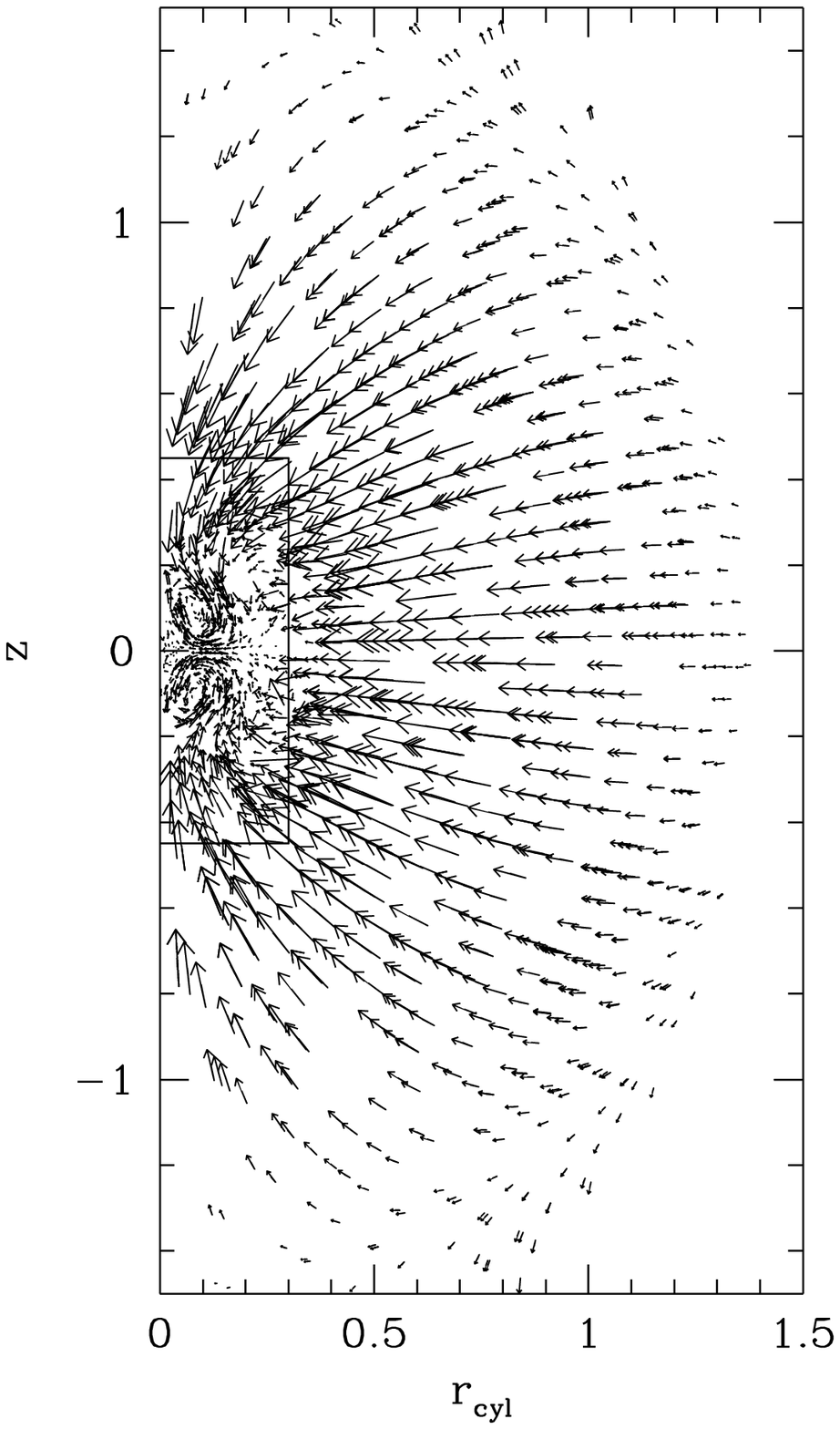} {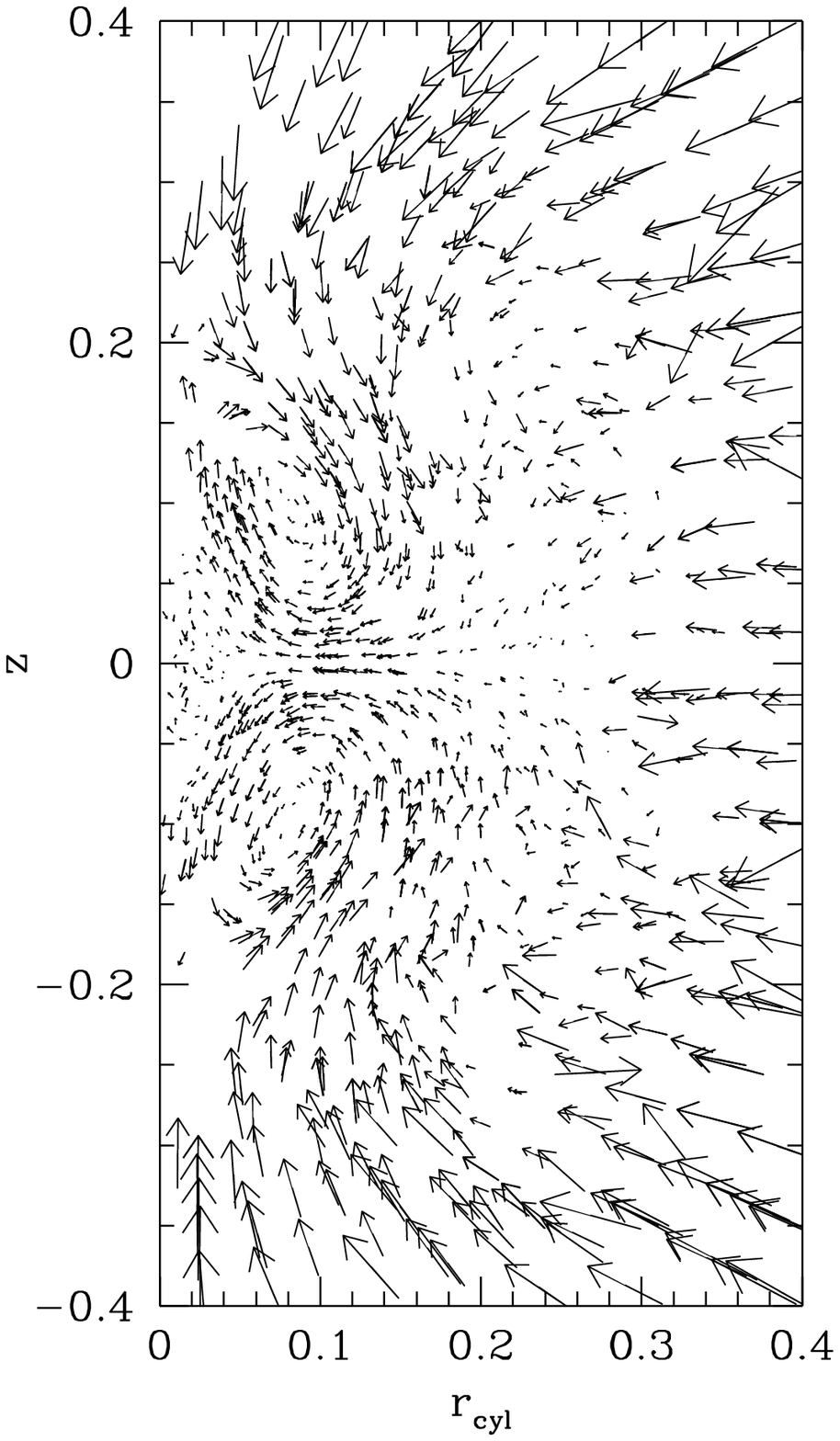}
\caption{Velocity field for a prolate protogalaxy ($q=1.25$) as a
function of cylindrical radius $r$ and position along the symmetry
axis $z$.  (a) The entire protogalaxy.  (b) Inner region of the
protogalaxy corresponding to the rectangular box in (a).}
\end{figure}

\begin{figure}
\epsscale{1.0}
\plottwo{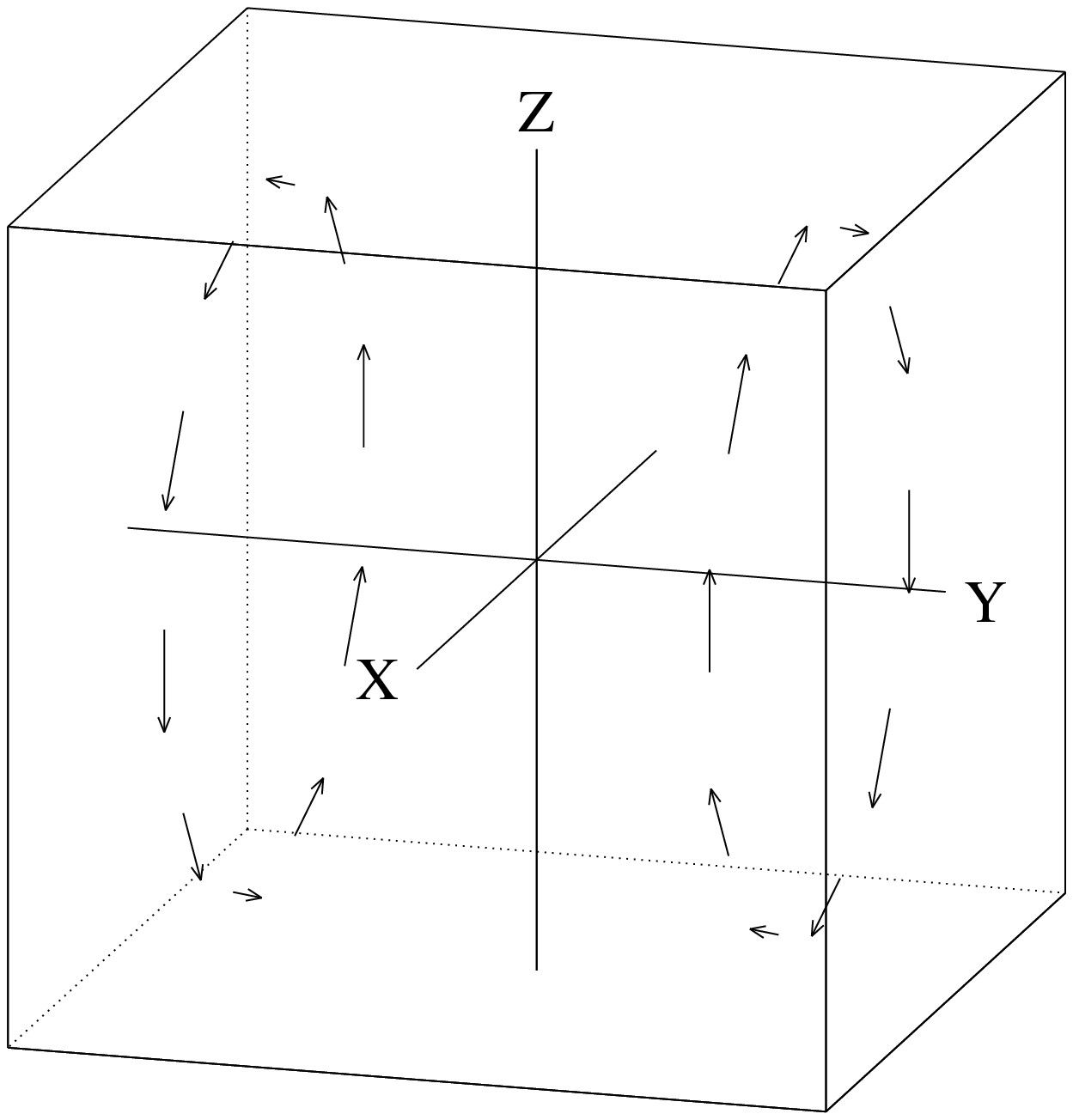} {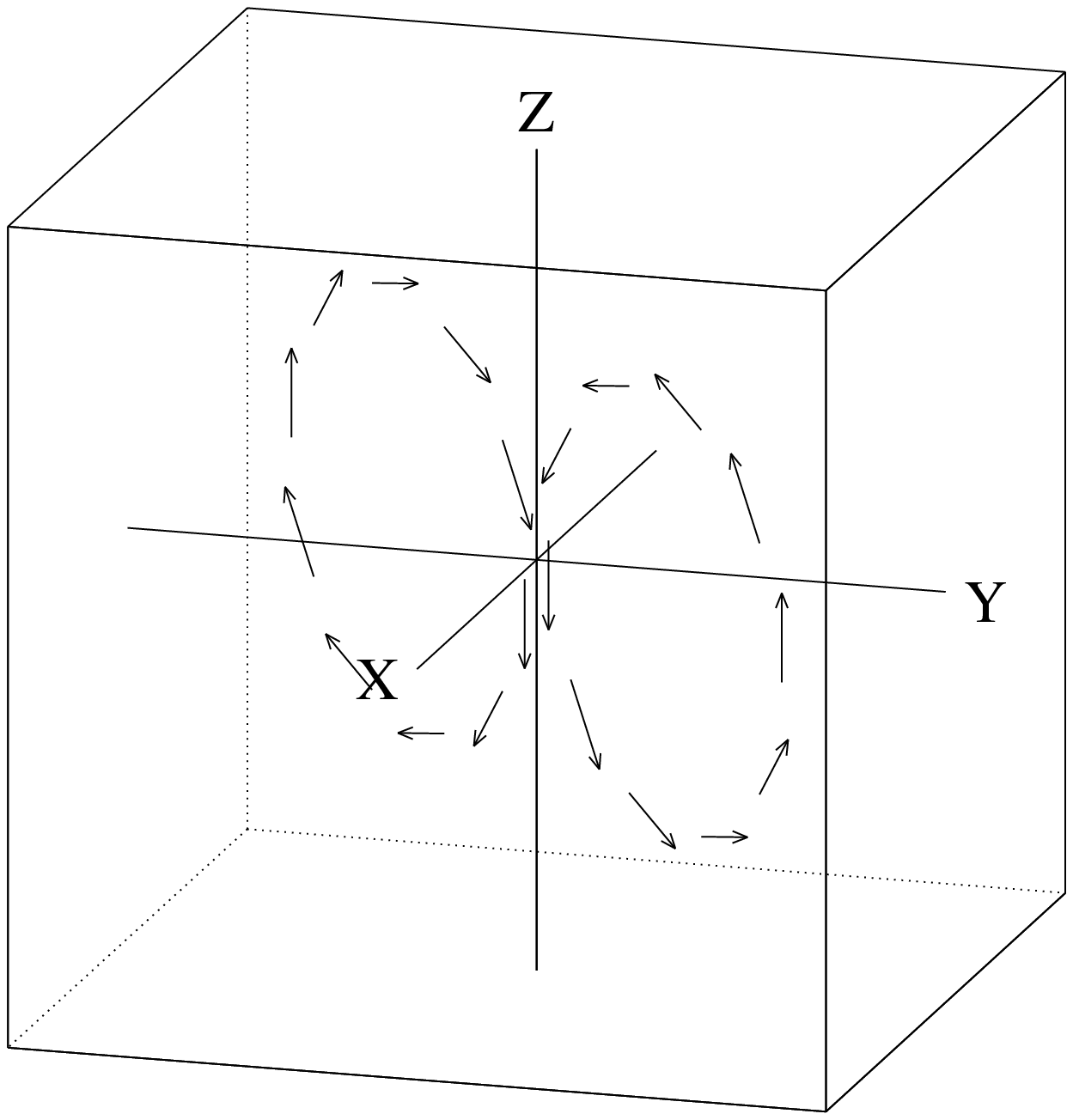}
\caption{Magnetic field in the Lagrangian
calculation described in Section 4.  (a) The initial field
configuration.  The symmetry axis of the protogalaxy coincides with
the y-axis.  (b) Final configuration.  Particles have evolved under
the influence of a shear field that has net angular momentum in the z
direction.}
\end{figure}

\end{document}